\documentclass[12pt,preprint]{aastex}

\begin{document}


\title  {Bayesian Single-Epoch Photometric Classification of Supernovae}

\author{Dovi Poznanski and Dan Maoz}
\affil{School of Physics \& Astronomy, Tel-Aviv University, Tel-Aviv
69978, Israel}
\email{dovip,dani @wise.tau.ac.il}

\and
\author{Avishay Gal-Yam\altaffilmark{1}}
\affil{Astronomy Department, MS 105-24, California Institute of Technology, Pasadena, CA 91125}
\email{avishay@astro.caltech.edu}

\altaffiltext{1}{Hubble Fellow.}

\begin{abstract}
Ongoing supernova (SN) surveys find hundreds of candidates, that require confirmation for their various uses: as
standard candles; to measure the cosmic supernova and star formation rates; to constrain progenitor models; and to track the contribution of supernovae (SNe) to metal enrichment. Traditional classification based on followup spectroscopy of all candidates is virtually impossible for these large samples.
The use of type-Ia SNe as standard candles is at an evolved stage that requires pure, uncontaminated samples.
However, other SN survey applications could benefit from a classification of SNe on a statistical basis, rather than case by case. With this objective in mind,
we have developed the SN-ABC, an automatic Bayesian classifying algorithm for supernovae. We rely solely on single-epoch multiband photometry and host-galaxy (photometric) redshift information
to sort SN candidates into the two major types, Ia and core-collapse supernovae. We test the SN-ABC performance on published samples of SNe from the SNLS and GOODS projects that have both broad-band photometry and
spectroscopic classification
(so the true type is known). The SN-ABC correctly classifies up to 97\% (85\%) of the type Ia (II-P) SNe in SNLS, and similar fractions of the GOODS SNe, depending on photometric redshift quality. Using simulations with
large artificial samples, we find similarly high success fractions for type Ia and II-P, and reasonable ($\sim75\%$)
success rates in classifying type Ibc SNe as core-collapse. Type IIn SNe, however, are often misclassified as Ia's. In deep surveys, SNe~Ia are best classified at redshifts $z \gtrsim 0.6$, or when near maximum.
Core-collapse SNe
(other than type IIn) are best recognized several weeks after maximum, or at $z \lesssim 0.6$.
Assuming the SNe are young, as would be the case for rolling surveys, the success fractions
improve, by a degree dependent on the type and redshift.
The fractional contamination of a single-epoch photometrically selected sample
of SNe Ia by core-collapse SNe, varies between less than 10\% and as much as
30\%, depending on the intrinsic fraction and redshift distribution of the core-collapse SNe in a given survey.
The SN-ABC also allows the rejection of SN ``impostors" such as active galactic nuclei (AGNs), with half of
the AGNs we simulate rejected by the algorithm.
SNe may be classified using just one or two bands, but three bands are necessary for obtaining a reasonable reliability and AGN rejection. Observations in a fourth band may not be worth the added observational cost, for classification purposes.
Our algorithm also supplies a good measure of the quality of the classification, which is valuable for error estimation.
\end{abstract}


\section{Introduction}

Supernovae (SNe) are usually subdivided into different types, based on spectroscopic or photometric criteria
\citep[e.g.,][]{FILIPPENKO_97}. The primary physical distinction among the common supernova (SN) types separates core collapse SNe (CC-SNe), that originate from the collapse of a massive star \citep[e.g.,][]{WOOSLEY_CCSNe}, from type-Ia SNe (SNe~Ia), for which there are competing scenarios, generally involving a thermonuclear runaway in a degenerate system \citep[e.g.,][]{Hilleb_IaSNe}.\\
Recent technological advances in wide field imaging and in automation capabilities, combined with a growing interest in transient phenomena, have led to numerous surveys that find hundreds of SNe. These surveys
search for SNe nearby \citep{Filippenko_01,Wood-Vasey_04}, at intermediate redshifts \citep{DILDAY_05,Cappellaro_05}, and at high redshifts \citep[][]{P07b,ASTIER_SNLS06,RIESS_04,Sollerman_05}.
While most ongoing SN searches focus on the use of SNe~Ia as standard candles for cosmological applications
\citep{RIESS_04,ASTIER_SNLS06}, large SN samples can be used also to study the statistical properties of the SNe themselves, and to measure their rates as a function of cosmic time \citep[e.g.,][]{MAOZ_AGY_04,DAHLEN_SNR04}.
Because the lifetimes of massive stars are short compared to cosmological timescales, the rate of CC-SNe is expected to follow the star formation history, and thus could provide an independent measure of this function
\citep[e.g.,][]{dahlen_99}. The rate of SNe~Ia is linked to star formation through a more complicated evolutionary process that depends on the nature of the progenitor systems. Measurement of the SN Ia rate may therefore constrain possible progenitor models \citep[e.g.,][]{Madau_98,AGY_MAOZ_04,STROLGER_04,Scannapieco_05,Mannucci_05,Mannucci_06,sullivan_06,forster_06}.

Future projects such as Pan-STARRS \citep{Kaiser_05}, LSST \citep{Stubbs_04}, and SNAP \citep{Aldering_05}, promise to increase the yield of SNe by orders of magnitude, especially at intermediate and high redshifts.
All of the SN studies above require large, well-defined samples of SNe with determined types and redshifts.
The common methodology of attempting to obtain spectroscopic followup of every SN candidate, in order to determine its type, its redshift (based on the host spectrum or from the SN spectrum itself), and possibly other characteristics, is unsuitable for this plethora of SNe, due to their faintness, sheer numbers, or both. For this reason, current surveys usually focus on subsets of the SN candidates for which spectroscopic observations can be obtained. Ongoing surveys already employ some photometric criteria for prioritizing followup, but
usually not for definitive classification, as we discuss below.

In \citet[][]{POZ_TP1} and \citet{AGY_TP2}, we showed, using color-color diagrams, that given a known redshift,
different types of SNe can often be distinguished based on colors alone. Our method
and webtool\footnote{\texttt{http://wise-obs.tau.ac.il/$\sim$dovip/typing/}} have been used, among others, by \citet{RAJ_TP}, to classify nearby SNe, by \citet{AGY_CCCP} to choose objects for followup prior to spectroscopy, and by \citet{STROLGER_04} for high redshift CC-SN
identification. The approach has been developed further by several studies. \citet{RIESS_TP} showed that, at high redshift, it might be even easier to distinguish, near maximum light, between SNe~Ia and CC-SNe, by virtue of the difference in the rest frame UV emission between these two classes of objects. \citet{JOHNSON_06}  investigated further the color-magnitude based classification of intermediate redshift SNe with different filter sets and both with and without redshift information. \citet{SULL_TP} introduced the use of automatic photometric classification as part of the reduction pipeline of the SN Legacy Survey \citep[SNLS;][]{ASTIER_SNLS06}, using two to three early light-curve points. The objective was mainly to recognize SNe~Ia, with a minimum of false positives and selection biases, and thus to prioritize targets for followup spectroscopy.
\citet{BARRIS_SNR06} fit complete light curves and applied various quality cuts in order to recognize SNe~Ia among the various transients they found in their survey \citep[see also][]{BARRIS_04}.

In this paper, we present an algorithm, the SN Automatic Bayesian Classifier (SN-ABC), intended for automatic
classification of SNe using single-epoch photometry and whatever redshift information is available.
The main novelties of our algorithm lie in the probabilistic and Bayesian methodologies.
Our approach is probabilistic in the sense that we do not expect to succeed with the classification
of every specific object, but to correctly classify most objects, while minimizing biases in the output sample,
this by assigning each object its most likely type.
The flexibility of the Bayesian framework allows us
to do so by analyzing SNe from different surveys, with different depths and types of redshift information,
to incorporate fully all the available information on each object, and to propagate correctly the unknowns. As a proof of concept, we focus on single-epoch photometry in three bands, but the method can be straightforwardly implemented with multiply sampled light curves with other photometric setups.
As we will show, SNe can be classified quite reliably even with meager input data, and therefore numerous SNe can be found and analyzed at a relatively low cost in telescope time.

We test our method on light curves, recently released by the SNLS, of 71 SNe~Ia and five ``plateau'' type (II-P) SNe \citep{ASTIER_SNLS06,NUGENT_IIP06},
and on the sample of 42 SNe found during the GOODS campaign \citep{DAHLEN_SNR04}. A third sample of simulated SN data is used
to investigate potential biases and the efficiency of the method as a function of SN and sample parameters. In a separate paper (Poznanski et al. 2007), we use our method to classify a sample of SNe we have found in the Subaru Deep Field \citep[SDF; ][]{Kashikawa_SDF}, with the objective of measuring the rates of Ia and CC SNe at high redshift based on this dataset.

Shortly prior to submission of this paper, \citet{kuznetsova_06} presented a Bayesian classification algorithm, and its application to SNLS and GOODS data. Although there are some similarities between their approach and ours, there
are also some major differences. First, while they use complete light curves, and discuss the importance of such data, we focus on single epoch observations, where information is scarce, and classification is less obvious. Second, they treat
the redshift of the SN as a known quantity without any ambiguity. While this is true for most of the SNe in the training samples they (and we) use, this is certainly not the case for existing and future SN samples, where methods of photometric classification apply. In our work, while we focus on the case where photometric redshifts for the host galaxies of the SNe are available, we treat the general case, in which each object can have any possible redshift estimation, either precise (spectroscopic) or totally unavailable. In each case the uncertainties in the redshift
measurement are taken into account. Finally, \citet{kuznetsova_06} estimate the probabilities after marginalizing over three extinction scenarios, (no extinction, $R_V=3.1$ with $A_V=0.4~$, and $R_V=2.1$ with $A_V=0.4~$). As explained in \S\ref{model}, we assume a range of extinction models with $R_V=3.1$, and $A_V$ between zero and one.

\section{Method}\label{method}
Our general approach for SN classification relies on a Bayesian template fitting technique.
Given a ``prior'' on the redshift of a SN candidate, we fit a set of template spectral energy distributions (SEDs) to the observed photometry of the SN. We then maximize the likelihood
in order to find the characteristics of the SN, derive the evidence for the object being either a SN Ia or a CC-SN, and find its posterior redshift distribution function. We describe below each step in detail.

\subsection{SN Photometry and Prior (Redshift) Information}\label{phot}

We will assume the availability of  single-epoch multi-band photometry of the candidate SNe. As we will show, this is probably the minimum amount of data that can produce scientifically meaningful results,
and is also the kind of information we have for the SDF (Poznanski et al. 2007). However, this method can be
easily applied to cases where more information is available, including
complete multi-band light curves, with a guaranteed improvement in the results.

We further assume the existence of a prior redshift probability distribution function (z-pdf) for the SNe.
Such a prior distribution would come naturally from the photometric redshifts of the host galaxies, but in principle could be from any source. Photometric redshifts are now being produced by every major extragalactic survey, and for well studied fields there may even be an available spectrum of the host galaxy, or one can be obtained later when the SN has faded. (In the Bayesian formulation, a spectroscopic redshift corresponds to a sharply peaked prior z-pdf.) In cases where there is no redshift information (e.g., because the host is too faint) one can assume a flat, or some other physically motivated, z-pdf.

The Bayesian approach permits a careful treatment of the unknowns, that can be propagated fully from one step of the analysis to the next. Even when the redshift is derived from spectroscopy, it can have uncertainties which are hard to propagate further in any frequentist approach. Here, we can treat self-consistently redshift information of any given quality, whether a narrow peaked z-pdf from a spectral redshift, a broad, or oddly shaped redshift distribution, including redshift limits, or even non-existent photometric redshift, all jointly analyzed. Furthermore, when interested in the redshift distribution of the SNe, one can use the full posterior
redshift distribution for each object, rather than a single value derived from its peak, and thus propagate further the complete information. Finally, in ``rolling searches'', where, e.g., all SNe are found when rising or near peak, one can similarly include a prior on the age of the SN, and thus improve the classification.

\subsection{Model}\label{model}

We use SN templates from the updated version of the SN SEDs of
 \citet[][]{NUGENT_02}\footnote{available at \texttt{http://supernova.lbl.gov/$\sim$nugent/nugent\_templates.html}}. These include sets of redshift $z=0$ spectra of SNe of different types and at different epochs, covering the NIR to the restframe UV.
After experimenting with various simulations and applications to real data, most of which are presented in Sections \ref{real} and \ref{MC}, we find that the best classification, and the fewest subsequent biases, in type and redshift distributions, are achieved when only the templates for types II-P and normal (stretch $s=1$) Ia are used in the fits.
As we will show in \S \ref{MC}, we find that, with the exception of type IIn SNe (see discussion in \S \ref{MC}), most CC-SNe, are more similar, on average, to type~II-P than to type Ia SNe. Since our main objective will be an empirical division between the two main physical classes of SNe, Ia  and CC, we retain only these two types, using the II-P templates as proxies for the whole CC SN class.
In general template fitting algorithms, the use of many templates, including relatively rare objects, tends to damage the fit of the more common objects \citep[e.g.,][]{Benitez_00}. In our case, adding templates for other CC-SNe, while somehow improving the classification of these types of SNe,  causes many SNe~Ia to be wrongly classified, and thus lowers the overall success rate of our algorithm.
We emphasize that while we use those two specific template families to measure whether a given object is a SN~Ia or a
CC-SN, we are concerned with the classification of \textit{all} types of SNe. Our working assumption (that is
largely confirmed by the results we present in the following sections) is that most type Ia SNe have colors and magnitudes more similar to those of a normal SN Ia than to those of a type II-P SN, and conversely, that most CC-SNe, including the various subtypes, are more similar to a II-P than to a SN~Ia.

Our adopted templates span a range in SN age, $t$, from about two weeks prior to maximum $B$-band light to two (three)  months past maximum for the Ia (CC) SNe. Using synthetic photometry, we have scaled the template spectra to reach at maximum light the $B$-band absolute magnitudes in Table 1 of \citet{DAHLEN_SNR04},
which are based on the works of  \citet{RICHARDSON_02}, \citet{LI_01}, and \citet{TONRY_03}. We consider two types of intrinsic dispersion in SN properties -- one in the absolute magnitude, which we take from \citet{DAHLEN_SNR04}, and one in color, of $0.1~\textrm{mag}$ (for whatever color is being considered, e.g., $r-i$, $i-z$, etc.). The latter is consistent with what has been measured for SNe~Ia by \citet{NOBILI_03}, \citet{JHA_06}, and Ellis et al. (in preparation), while for CC-SNe we are not aware of the existence of similar published data \citep[though see][]{Elmhamdi_03}.

We apply to the template spectra a grid of redshifts (from 0 to 2) and extinctions ($A_V=0-1~\textrm{mag}$) at the SN redshift, using the \citet*{CCM} extinction law, with $R_V=3.1$. Luminosity distances are calculated
using the currently favored cosmology ($h=0.7,\Omega_\Lambda= 0.7,\Omega_m= 0.3$).
Finally, we apply a range of observer-frame Galactic extinctions to the spectra (from zero extinction to $A_V=0.1$,
the maximal value for the SNe in the GOODS and SNLS surveys), again using the reddening law of \citet{CCM}.

We calculate synthetic magnitudes from the SEDs, based on the bandpasses used in the particular survey being analyzed. These bandpasses are the products of the filter transmission curves, detector quantum efficiency curves, and (for ground-based data) atmospheric
transmission. We thus obtain a $5d$ data cube of model magnitudes for each type of SN as a function of redshift, age, band, host extinction, and Galactic extinction, from which we select for each object a $4d$ cube according to the expected Galactic extinction from the maps of \citet*{SFD}. We perform all of our analysis using Vega-based magnitudes.

\subsection{Fitting}\label{fitting}

For every SN we classify, we calculate two $3d$ posterior likelihood matrices, one assuming it is a SN Ia, and one assuming it is a CC-SN:
\begin{displaymath}
L_{type}(z,A_V,t)=P(z) \cdot \mathrm{exp}\left[-\sum_i{\left(\frac{m_{\mathrm{mod},i}(z,A_V,t)-m_i}{dm_i}\right)^2}\right],
\end{displaymath}
with the summation being over the available bands, and denoted here by the index $i$. $P(z)$ is the redshift prior,  $m_{\mathrm{mod},i}$ are the model magnitudes computed from the templates (see \S \ref{model}), $m_i$ are the measured magnitudes, and $dm_i$ are the photometric errors and model dispersions summed in quadrature. In order to account for the different dispersions in the absolute magnitudes of the model SNe and their colors, we technically do not fit $n$ magnitudes in $n$ bands, but rather one magnitude and $(n-1)$ colors.
We marginalize the likelihood function over the three parameters, $z$, $A_V$, and $t$, and obtain the summed likelihood for each type of SN, commonly referred to in Bayesian literature \citep[e.g., ][]{GELMAN_BAYS} as the \textit{evidence},
\begin{displaymath}
E_{type}=\int {L_{type}(z,A_V,t)~dz~dA_V~dt}.
\end{displaymath}

In order to classify the SNe, we define a relative evidence to express the probability that an object is a SN Ia, as
\begin{displaymath}
P_{Ia}=\frac{E_{Ia}}{E_{Ia}+E_{CC}}.
\end{displaymath}
We also derive for each SN its posterior redshift, assuming it belongs to a given type of SN, by marginalizing the relevant likelihood function over the nuisance parameters, namely the age and the extinction.
The final output is then the probability for the SN to be of type Ia, rather than being a core-collapse event, and its posterior z-pdf. In the following sections, we consider the adopted type of a SN to be the type which has a probability higher than $0.5$. Naturally, the higher this probability, the more secure is the classification. We also examine the $\chi^2$ value for the best fitting template, in order to ascertain the goodness of fit in an absolute sense, and in order to reject SN impostors, mainly active galactic nuclei (AGNs).

\section{Tests on Real Data}\label{real}

We now apply our algorithm to two real data sets, SNLS and GOODS, that have photometric data of the type we are considering (e.g., SNe observed in at least three bands), and followup spectroscopy based upon which we can evaluate the performance of our method.

\subsection{SNLS}\label{SNLS}

\subsubsection{Ia SNe}\label{SNLS-Ia}
We begin with 71 published lightcurves of SNe~Ia from the SNLS project\footnote{ available at \texttt{http://snls.in2p3.fr/conf/release/}}, described in
\citet[][]{ASTIER_SNLS06}. The SNLS is a 5-year `rolling' SN-Ia survey, using the MegaCam wide field imager on the 3.6 m Canada-France-Hawaii Telescope. The fields are imaged in four bands, similar to the Sloan Digital Sky Survey \textit{g,r,i,z} \citep{FUKUGITA_96,ASTIER_SNLS06}, with limiting magnitudes in $i$ of the order of $24.5~\textrm{mag}$. The SN candidates are confirmed using spectroscopy from 8-10 m class telescopes.
The SNe in our sample are all spectroscopically confirmed SNe~Ia at $z=0.25-1$, with a median of $z=0.6$. Typical photometric errors are smaller than $0.1~\textrm{mag}$. Since our main focus is on classification by means of three bands, we ignore at first
\textit{g}-band data (but see \S\ref{bands}), and use for each SN only those data combinations which
have same-day ($\pm1$) photometry in \textit{r,i,z}. This produces a sample of 172 `objects', extracted from 58 SNe. This selection does not affect the redshift distribution, which remains similar to the distribution of the original sample.

Since these SNe have spectroscopic redshifts, we create simulated photometric redshifts, $z_{\mathrm{rand}}$, by adding to each object's redshift a normally distributed (with zero mean and $\sigma_1$ standard deviation) error,
\begin{displaymath}
z_{\mathrm{rand}}=z_{\mathrm{spec}}+\mathrm{noise}(\sigma_1),
\end{displaymath}
and create a Gaussian z-pdf of width $\sigma_2$, centered on the noise-added redshift,
\begin{displaymath}
P_{\mathrm{phot}}(z)=\frac{1}{\sigma_2 \times \sqrt{2 \pi}} \cdot \mathrm{exp}\left[-\frac{(z-z_{\mathrm{rand}})^2}{2\sigma_2^2}\right] .
\end{displaymath}
To evaluate the impact of the precision of redshift determination, we have run the SN-ABC for various combinations of $\sigma_1$ and $\sigma_2$ values. For every SN, we perform several specific $z_{\mathrm{rand}}$ realizations, and average the outcomes.

\begin{figure}
\epsscale{.98}
\plotone{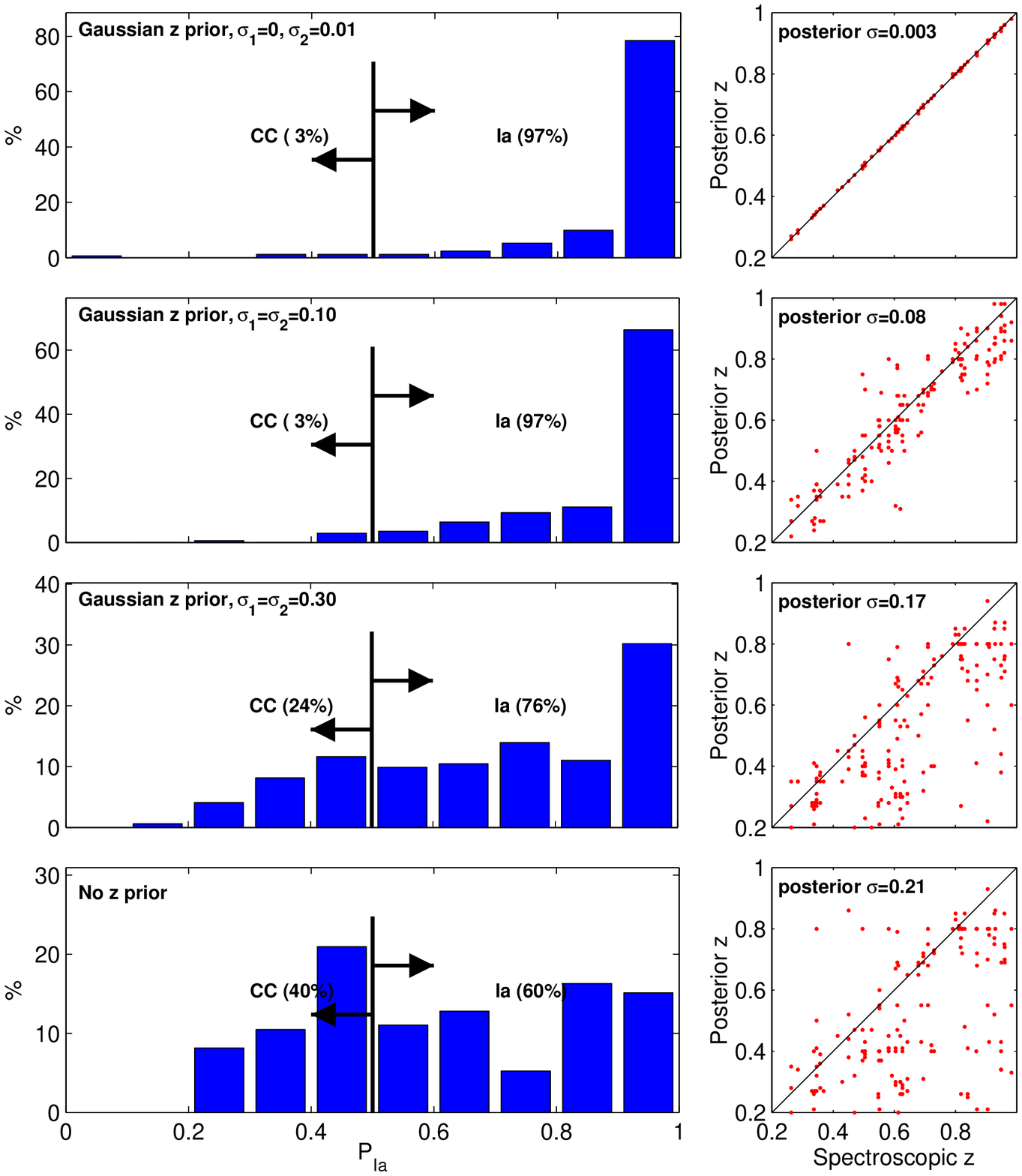}
\caption{Classification of SNe~Ia from SNLS. \textit{Left:} distribution of the type determination parameter $P_{Ia}$ for different redshift priors.
Values of $P_{Ia}$ higher than $0.5$ mean the object is probably a SN~Ia, while smaller than $0.5$ mean it is a CC-SN.
\textit{Right:} posterior redshifts for the priors on the left. Solid line is $z_{\mathrm{posterior}}=z_{\mathrm{spec}}$, for reference. The SN-ABC works well, assuming high quality photo-z values are available (upper two panels).\label{SNLSIa}}
\end{figure}

In Figure \ref{SNLSIa} we present the SN-ABC output $P_{Ia}$ distributions (left) for various precisions
of redshift determinations, and the posterior redshift compared to the spectroscopic one (right).
We start with $\sigma_1=0$ and $\sigma_2=0.01$, which can represent well-determined, i.e., spectroscopic, redshifts. In this case $\sim97\%$ of the SNe are correctly classified by the SN-ABC.
With $\sigma_1=\sigma_2=0.1$, a reasonable precision for photometric redshifts in most surveys and redshift ranges \citep[e.g.,][]{Grazian_06,COE_06}, this result remains unchanged, with
$97\%$ correctly classified SNe, and $\sim65\%$ having $P_{Ia}>0.9$. The posterior redshift scatter is marginally reduced, compared to the constructed prior standard deviation of $\sigma=0.1$, indicating that this precision in redshift is about the limit of what can be achieved with single epoch data in three bands.

For $\sigma_1=\sigma_2=0.3$, i.e., very broadly distributed and imprecise redshifts, and probably the worst case scenario for photometric redshifts, we correctly classify $\sim76\%$ of the SNe as type Ia, and improve the scatter in the redshift determination from $\sigma=0.3$ to an \textit{a posteriori} value of $\sigma\sim0.17$, thus improving the redshift dispersion by almost a factor of 2. On the other hand, it can clearly be seen that the posterior redshift distribution is heavily biased towards lower $z$. This is a consequence of the fact that, the lower the redshift, the more freedom the minimization has in the other parameters. At lower $z$ the SN template is inherently brighter and can be `aged' and extinguished in order to fit the observed magnitudes, while at higher $z$ it must be closer to peak and less extinguished.

The most extreme scenario is the absence of any prior on the redshift. This is probably relevant only to the few objects that have no measured host due to its faintness. In this case, we drop to only $\sim60\%$ correct classifications, only slightly better than random assignment of SN type. Not surprisingly, parameter space is wide enough to accommodate both types, when no information on the host redshift exists, and the SN light/color curve is so scantily sampled.

We can thus conclude that single epoch photometry in three bands, combined with a reasonably well determined host galaxy redshift, is sufficient to recognize SNe Ia, with only a few false negatives, and high confidence levels.

\subsubsection{II-P SNe}
In order to measure the performance of the SN-ABC in correctly classifying CC-SNe,
we repeat the procedure described in \S \ref{SNLS-Ia}, applying the same ``pseudo photometric redshifts''
to the much smaller sample of five type II-P SNe, presented in \citet{NUGENT_IIP06}. We extract 25 `objects' with same-day photometry from among three of these SNe, which are at redshifts of 0.13 to 0.21.

\begin{figure}
\epsscale{.8}
\plotone{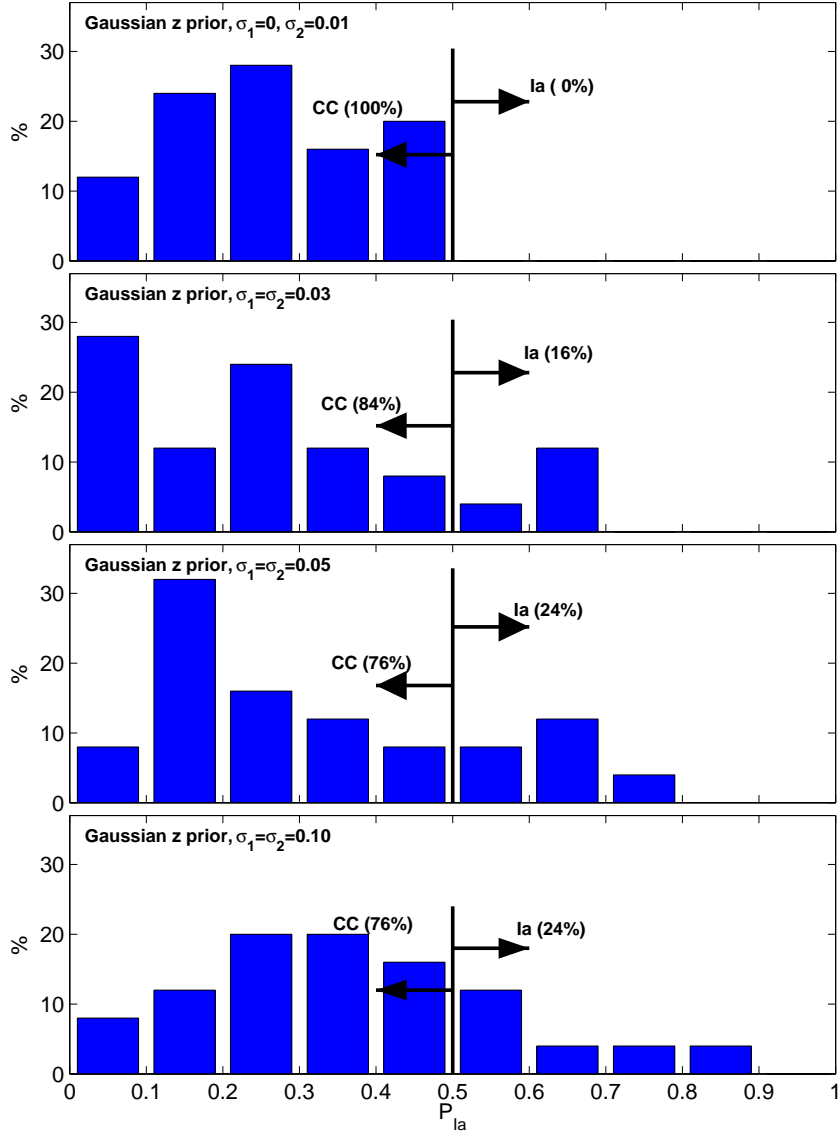}
\caption{Same as Figure \ref{SNLSIa}, for SNLS SNe II-P with four different redshift priors.\label{SNLSIIP}}
\end{figure}

As shown in Figure \ref{SNLSIIP}, when using the precise spectral redshifts, we achieve a perfect success rate with not a single object misclassified. When using broader, more realistic, z-pdfs, we reach success rates between $85\%$ ($\sigma_1=\sigma_2=0.03$) and $75\%$ ($\sigma_1=\sigma_2=0.1$). Since the redshifts of these SNe are significantly lower than those of the SNLS Ia sample, we examine cases with comparatively smaller values of $\sigma_1$ and $\sigma_2$ that are more reasonable for photometric redshifts in this range.

In all the simulated photo-$z$ schemes we have examined, both for the Ia and CC-SNe, the minimum $\chi^2$ values are lower than the value that defines the $99\%$ confidence interval on the goodness of fit, which we derive from simulations, as will be described in \S \ref{MC}, below. The best fit templates therefore fit the data well in an absolute sense.

While the tests on the SNLS samples, both type Ia and II-P, indicate that our algorithm works well, these results may be misleading since these samples were selected by the SNLS team for their quality, and are not in any sense a complete representation of all the SNe found during the survey. The SNLS pipeline includes a color-based pre-selection of objects for spectroscopy, and the sample might be biased against SNe~Ia with uncharacteristic colors.
In order to check our method
on a complete sample, we apply the SN-ABC to all of the SNe found during the \textit{Hubble Space Telescope} (HST) GOODS campaign.


\subsection{GOODS}\label{GOODS}

We proceed with a SN sample from the GOODS SN survey. As part of this project, two fields of about $150~ \textrm{arcmin}^2$ each were observed with the ACS camera on HST every 45 days, over five epochs, in order to search for SNe. We have compiled from \citet{RIESS_04} and \citet{STROLGER_04}, all available photometry for the 42 SNe found in the GOODS fields. These SNe cover a redshift range of $0.2$ to $1.55$, with a median of $0.76$, which is higher than the SNLS range. The light-curve coverage is more limited than for the SNLS sample, and not all SNe have spectroscopic redshifts or types. However, GOODS  is a complete sample, without any subsequent selection. The SNe were classified following a decision scheme presented in \citet{STROLGER_04}, and the whole sample was divided into three subsets -- ``gold", ``silver", and ``bronze" -- according to the reliability of the type determinations.

As in \S \ref{SNLS},
we first reject all the SNe which do not have same day ($\pm1$) photometry in three bands (in this case F606W, F775W, \& F850LP), leaving 28 SNe. After excluding one SN with no redshift information, we extract
from the remaining 27 SNe (13 Ia, 14 CC), 41 same-epoch photometry ``objects'', of which 23 are of type Ia, and 18 are CC. Among these 41 objects, 28 (17 Ia, 11 CC) have spectroscopic redshifts while 13 (6 Ia, 7 CC) have only photometric redshifts. The requirement of same-epoch photometry changes the redshift distribution of the sample, mainly by rejecting the few SNe with $z>1$, and leaves a distribution with a median redshift of $\sim0.6$, ranging from 0.2 to 1, similar to the SNLS Ia sample.

For those SNe with spectroscopic redshifts we create pseudo-photo-$z$'s, repeating the approach in \S \ref{SNLS}, with $\sigma_1=\sigma_2=0.1$. When applied to this sample, the SN-ABC correctly classifies 16 of the 17 SNe~Ia and 9 of the 11 CC-SNe.
The single SN~Ia which is misclassified is the last of three observations of SN 2002hr, a SN~Ia, about two months past maximum brightness. The previous two epochs of the same SN are well classified. The misclassified CC-SNe are SNe 2002kb and 2003bb.
For $\sigma_1=0$, $\sigma_2=0.01$, representing well determined
spectral redshifts, the success rates remain unchanged for the CC-SNe, but for the SNe~Ia one more object is misclassified, one of the three epochs of SN 2002ga, a bronze SN Ia. Since it is well classified with a broader z-pdf, its misclassification may be due to an underestimated redshift uncertainty.

For those SNe with photometric redshifts, we assume a Gaussian pdf with $\sigma_2=0.1$, since we do not have the real z-pdfs.
For this photo-$z$ sample, we recover the same classification as that of the GOODS team for only two of the six SNe~Ia and four of the seven CC-SNe. Increasing $\sigma_2$ to a value of $0.2$, to test whether we have underestimated the uncertainty in the redshift, does not improve this result.

It is clear that the SN-ABC functions extremely well on the GOODS SNe with spectroscopic redshifts, mostly in accord with the classification presented in \citet{STROLGER_04}, while for the SNe with photo-$z$'s there are severe discrepancies. This illustrates the difficulty to classify SNe without spectroscopy, and it is hard to ascertain which classification scheme is correct for every object.
If we consider agreement between the GOODS classification and ours as an indicator of the quality of the classification itself (either the SN-ABC or GOODS), it is clear that our division of the sample according to the existence or non existence of spectral redshift splits the sample into distinct quality classes. This is less true for the GOODS gold/silver/bronze quality flags (though one should note that these divisions are not independent). For example, our classification of the CC-SNe with spectral redshifts, belonging to the bronze sample, is in agreement with the GOODS type for all the objects, while it is in disagreement for one of the four bronze CC-SNe with photo-$z$.
More dramatically, for the silver CC-SNe, neither of the two SNe with photo-$z$ are similarly classified, while four out of the five with spectral redshifts are in agreement. Similar trends are seen with the type-Ia SNe.
Since all the SNe~Ia in the gold sample have spectral redshifts, this issue has no implications for this particular, higher quality, subset. We note that, by stating that the subsample with photometric redshifts is of lower quality, we do not imply that the problems arise necessarily from the photometric redshifts themselves, since we do not have the information needed to reveal the source of the discrepancies. The fact that the SN-ABC and GOODS classification for the SNe with spectra
are in excellent agreement, shows that our algorithm works well. The fact that there are significant discrepancies when
classifying SNe with no spectra indicates that the GOODS classification of at least some of these objects might be erroneous. This may, alternatively, indicate a weakness in the method. If photo-z's confidence limits do not represent faithfully the true uncertainty in redshift, then our algorithm could give incorrect results.
As in \S \ref{SNLS}, there are no objects with minimum $\chi^2$ values outside the $99\%$ confidence interval.

An advantage of our SN-ABC procedure is that it provides a quantitative indication of the quality of the classification, in the form of the $P_{Ia}$ value, that can be used for subsequent error estimation. The issues of classification and redshift uncertainties were addressed in \citet{RIESS_04} by deriving separate Hubble diagrams from the gold and silver samples (and ignoring the bronze sample).
\citet{DAHLEN_SNR04} deal with the uncertainties in the classification only (i.e., ignoring the possible errors in redshift determination), by testing how their results on SN rates change if the entire bronze sample
is misclassified. \citet{STROLGER_04} used the entire GOODS sample, irrespective of quality, in their derivation of the SN Ia time delay distribution. The last treatment does not test for possible biases
that arise from errors in classification or redshift determination.

\section{Tests on a simulated SN sample}\label{MC}

While the GOODS sample is complete, it is quite small and even smaller when restricted to SNe
with high-confidence classification. This motivates us to explore biases that might arise from our classification algorithm, and the dependence of the success rate on various parameters, by using a large simulated sample of SNe.
The simulated sample also permits us to test the performance of the algorithm when the input data are additional CC types, as well as SN ``impostors" such as AGNs.
In \S\ref{bands} we examine the influence on the classification of having more, or less, than three photometric bands.

We have randomly generated $5,000$ SNe of each of the SN types: Ia, Ibc, II-P, and IIn. To each
SN we assign a redshift, an epoch, and a host extinction. These characteristics are drawn from a uniform distribution. While these are not the true distributions, intrinsic or observed, they hold two advantages
for our objectives. First, flat distributions facilitate the discovery of possible trends in the reliability of the SN-ABC classification, by creating many SNe with every combination of properties. Second, the true distributions of SN rates as a function of redshift, type, extinction etc., are yet unknown, or highly uncertain. Assuming particular
distributions could introduce unknown systematics into the comparative success rates.
The results could also be used incorrectly to draw, for example, the overall contamination of a SN~Ia sample by CC-SNe. The contamination fraction, while very valuable for many SN searches,
depends strongly on the intrinsic rates of the different types of CC-SNe as a function of redshift.
Such an estimate of contamination must therefore be tailored to a specific survey, taking into consideration the range of possible distributions in SN properties.

For comparison of the results based on our simulated SNe to those based on the real SN samples, we mimic the observational properties of the GOODS sample with its three HST bandpasses, photometric errors and limiting magnitudes. The results in this section are for this specific configuration, and serve as an illustration of the capabilities of the SN-ABC.

As in \S \ref{model}, we calculate  the synthetic magnitudes of the fake SNe using the SEDs from the spectral templates of \citet{NUGENT_02} for types Ia, Ibc, \& II-P SNe. Since the \citet{NUGENT_02} spectra of type IIn SNe  are theoretical blackbody SEDs, for this type only we use the templates from \citet{POZ_TP1}. As in \S \ref{model},
absolute magnitudes and their dispersions are taken from \citet{DAHLEN_SNR04}. The scatter in color within each type
is applied to the CC-SNe by adding an intrinsic, normally distributed, noise with standard deviation of  $\sigma=0.2~\mathrm{mag}$, the value used by \citet{SULL_TP}. The SNe~Ia we simulate are given different stretches $s$, following the method described in \citet{SULL_TP}. We simulate a Gaussian distribution of stretches with an average of $s=1$ and a dispersion of $\sigma=0.25$, truncated in the range $0.6 \leq s \leq 1.4$. We model the stretch-luminosity relation using the formalism $M_{Bc}=M_B-\alpha(s-1)$ \citep{Perlmutter_99}, where $M_{Bc}$, and $M_{B}$ are the corrected and uncorrected $B$ band absolute brightnesses respectively, and the correlation factor is $\alpha=1.47$. We also apply color-stretch corrections using the method presented in \citet{KNOP_03}, by dividing the template spectra of normal, $s=1$, SNe~Ia by smooth spline functions, in order to match their restframe $UBVRI$ colors
to those of SNe with various stretches.
We have measured the mean photometric errors of the GOODS sample in each band as a function of magnitude, and again assuming the noise is normally distributed, have added it to each object. We further  assigned to
the SNe a pseudo-photo-$z$ with $\sigma_1=\sigma_2=0.1$ (see \S \ref{SNLS-Ia}).

\subsection{Type Ia SNe}

\begin{figure}
\epsscale{0.95}
\plotone{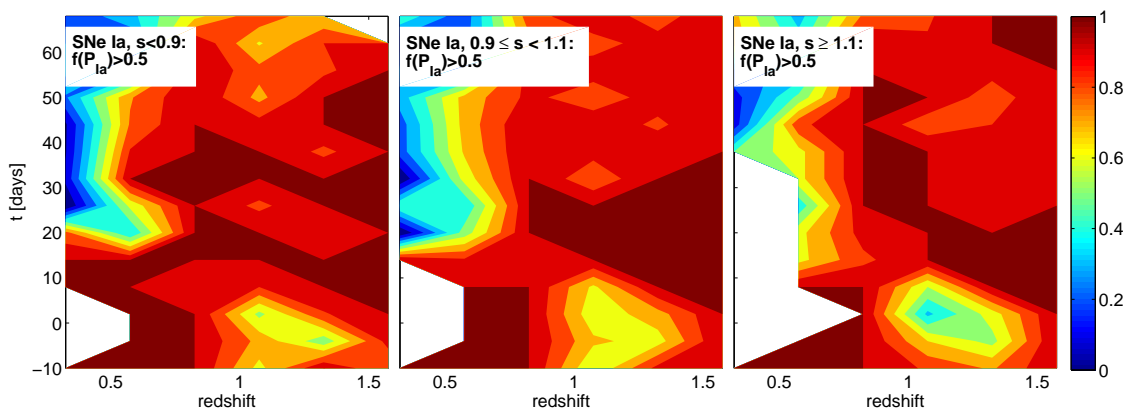}
\caption{Classification success rate for simulated type-Ia SNe with three stretch ranges, as marked,
as a function of redshift and age.
Contours trace $f(P_{Ia}>0.5)$, the fraction of SNe that are successfully classified (i.e., having $P_{Ia}>0.5$).
The solid curves show the 50\% success level, demarcating the
regions of relative success and failure of the method. Regions of parameter space where
we have no simulated objects (usually because they were too bright or faint to be in this
GOODS-like sample) appear in white. Note the similarity of the three plots indicating a weak dependence of classification success rate on the SN stretch/absolute magnitude.
\label{f:MCstr}}
\end{figure}

Figure \ref{f:MCstr} shows the average success rate of our algorithm, as a function of SN age and redshift, for
three different stretch ranges. Each panel shows in contours the fraction of SNe~Ia that are correctly classified by
the SN-ABC, i.e., the fraction of SNe that are given $P_{Ia}$ values higher than half. The solid curves show the $50\%$ success level,
demarcating the regions of relative success and failure of the method. The regions of parameter space where we have no simulated objects (usually because they were too bright or faint to be in this GOODS-like sample) appear in white.
As can be seen, the differences in plots for different stretch ranges are minor, showing that the diversity in color and magnitude among SNe~Ia does not affect the classification significantly. In the subsequent analysis, we marginalize over the full range of stretch values.

\begin{figure}
\epsscale{0.95}
\plotone{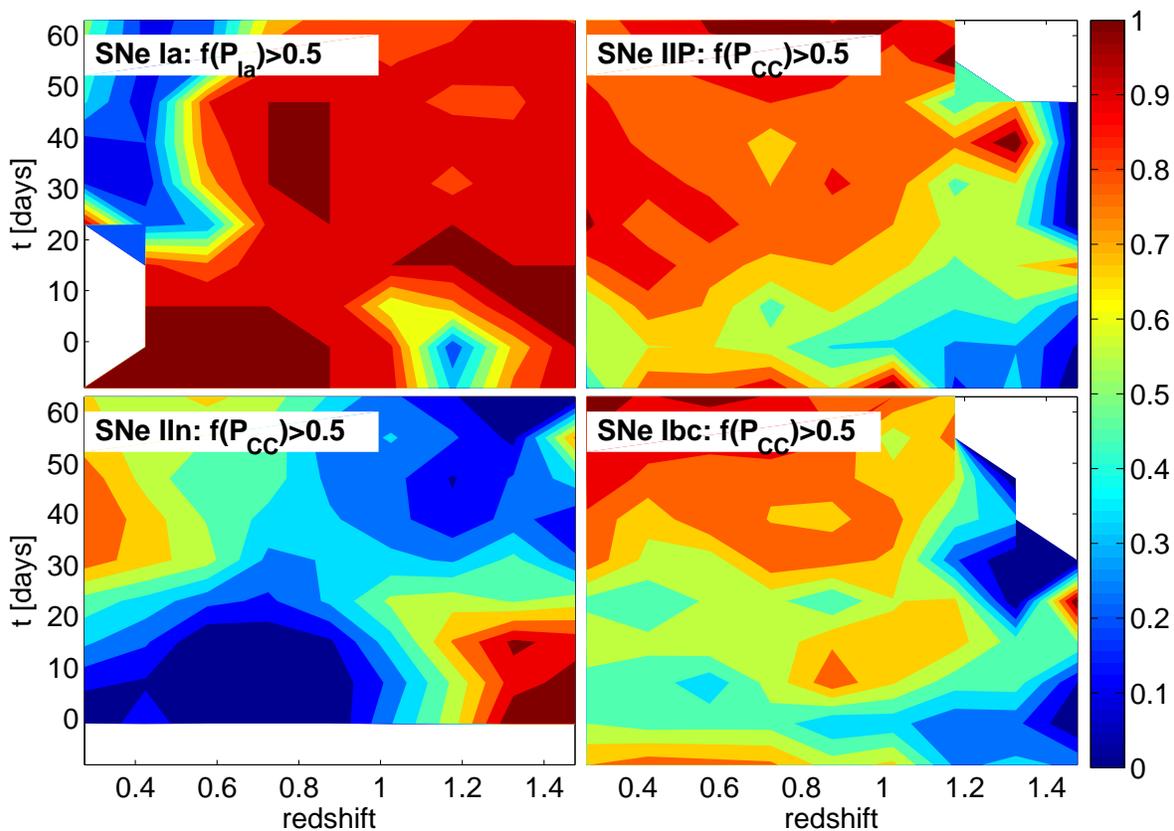}
\caption{Same as Figure \ref{f:MCstr}, but for simulated SNe of types Ia, II-P, Ibc, and IIn, as marked.
The top left panel is effectively the weighted sum of the three panels of Figure \ref{f:MCstr}. Types II-P and Ibc
are correctly classified as core-collapse SNe for a large range in age and redshift, but SNe IIn are generally
misclassified as SNe~Ia.
\label{f:MCscss}}
\end{figure}

\begin{figure}
\epsscale{0.95}
\plotone{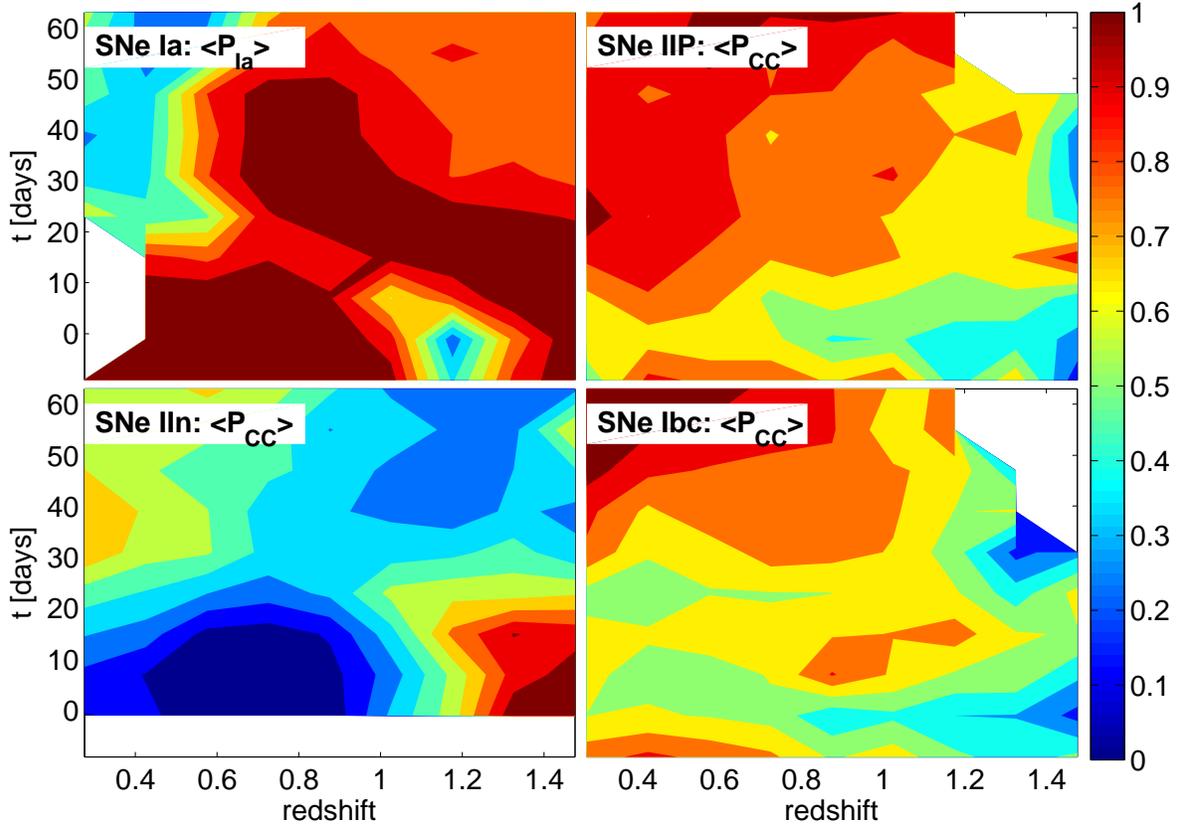}
\caption{Same as Figure \ref{f:MCscss}, but plotting contours of mean $P_{Ia}$ (top-left panel) or $P_{CC}$ (top-right and bottom panels) values for simulated SNe of types Ia, II-P, Ibc, and IIn, as marked.
The solid curves show the $P=0.5$ level.
Note the similarity with Figure \ref{f:MCscss}, indicating that the SN-ABC output, $P_{Ia/CC}$, is a good indicator of the reliability of the classification.\label{f:MCavP}}
\end{figure}

As can be seen in the top-left panel of Figure \ref{f:MCscss} (which is basically a weighted sum of the three panels of Figure \ref{f:MCstr}), the SN-ABC success rate on simulated SN~Ia data is high, greater than $90\%$,
in most regions of parameter space. The notable exception to this success is the population of 1-2 months old, relatively low-$z$, SNe. As already shown in \citet{POZ_TP1}, such SNe~Ia have blue rest-frame colors that are similar to those of CC-SNe. Furthermore, the older the SN, the fainter it is, and therefore it can be erroneously fitted more easily with younger CC-SN templates, which are intrinsically dimmer. This means that, by excluding objects from a sample below a certain $z$, one can obtain higher classification success rates.
The top-left panel of Figure \ref{f:MCavP} shows, for the same simulated SNe~Ia, the average $P_{Ia}$ value
(rather than the success rate) as a function of age and redshift. Clearly, most of the SNe are not only classified correctly but their classification has a high confidence, with an average $P_{Ia}$ greater than $0.8$ in most regions. We also note, from a comparison of the top-left panels of Figures \ref{f:MCscss} and \ref{f:MCavP} that the average $P_{Ia}$ value
follows quite well the actual success rate, meaning that it can serve as a reliable quality indicator.
We have searched for trends in the success rates of the classification as a function of all the other parameters, such as color, photometric errors, and extinction, and found no obvious correlations.

\subsection{Core-Collapse SNe}

In the three remaining panels of Figure \ref{f:MCscss},  we can see that the
picture is more complex within the zoo of CC-SNe. Most type-II-P SNe are correctly classified, being progressively better classified at later times. This is the complementary image of what we see with type Ia SNe. The colors of young II-P SNe resemble those of Ia SNe about a month past maximum light. For this reason type II-P SNe are also better classified at low~$z$.

The SN-ABC fails with type-IIn SNe, with about half of them being erroneously classified as SNe~Ia.
This is the result of a degeneracy in color-magnitude space between these two types, and for which
we have no straightforward solution. As already noted in \S \ref{model}, we do not use in the SN-ABC
a set of type IIn templates, because their addition, while improving the classification of SNe IIn as CC-SNe,
would considerably hinder SN~Ia classification. Type IIn SNe likely result from the core-collapse of very massive stars \citep{AGY_05gl}, the high-mass tail of the initial mass function, and
are therefore suppressed in volume limited samples. \citet{Cappellaro_97} measure their fraction to be $\sim2-5\%$ of the CC population in the local universe. Nonetheless, due to their brightness, in magnitude limited samples their fraction could be as high as $\sim15-20\%$.
For example, in IAU circulars between January 2005 and September 2006, the Nearby SN factory \citep{Wood-Vasey_04}, and the SDSS II SN survey \citep{DILDAY_05}, have together  reported  $\sim11$ type IIn SNe out of  $\sim55$ CC-SNe.
SNe IIn appear to be an important contaminant of future SN~Ia samples, and may cause difficulties for
cosmology oriented studies. This may have already affected some current studies that relied to some degree on photometric classification, e.g., \citet{BARRIS_SNR06}.
Such contamination could explain their measured SN rates, which seem inconsistent with other published results, as discussed by \citet{Neill_06}. The possible contamination of SN-Ia samples by type-IIn SNe has also been discussed by \citet{Germany_04}.

Our algorithm correctly classifies most type Ibc SNe as CC-SNe, but with lower success rates than type II-P SNe. Considering the fact that we are using II-P templates to recognize Ibc events, this should not be surprising. As with SNe II-P, when SNe~Ibc are near peak, their colors
are degenerate with those of SNe~Ia.
A caveat however, is that the templates from which we have simulated the type-Ibc SNe may not truly represent the high-$z$ population, a possibility that must await future spectroscopy of such events.
As was the case for SNe Ia, we find no trends in the SN-ABC results as a function of all the other simulated parameters, such as magnitudes in the different bands, photometric errors, and extinction.
The upper-right and two bottom panels of Figure \ref{f:MCavP} show that for CC-SNe, as was the case for SNe~Ia,
the average $P_{CC}$ values follow the success rate of the SN-ABC; when the success rate is low, the $P_{CC}$ values are, on the average, also lower, indicating when the fit is poor and thus serving as measures of quality for the classification of each object.

In order to simulate "rolling" surveys, where a field is imaged
 repeatedly, and the SNe are found when they are young,
we have run simulations using only SNe younger than three weeks past explosion.
Assuming a young SN age during classification, i.e., marginalizing the likelihood only over the relevant
extent in age, improves the success fractions of the SN-ABC significantly, as expected, since it removes
many of the young vs. old degeneracies. The improvement depends on the redshift range and SN type.
For example, for redshifts between 0.4 and 0.6, the success fraction for SNe~Ia improves from 91\% to 93

\subsection{Misclassification and Contamination by AGNs}
The fractional contamination by false positives (e.g., non-Ia's
among SNe classified as Ia) will depend on the intrinsic distribution
among types in a given survey, which depends on the depth of the survey.
For example, shallow, flux limited, SN surveys at $z\sim 0.1$ find, following
spectroscopic confirmation, an 80\% Ia fraction (Gal-Yam et al. 2007 in preparation).
Since this fraction is based on a small number of SNe, we will assume it could
be as low as 50\%.
We will also assume for simplicity that the CC-SNe are equally divided between the three
subtypes, II-P, IIn, and Ibc. The following numbers depend weakly on this assumption.
Averaging over CC-SN subtypes, about 30\% of the CC-SNe in this redshift
range ($z\sim 0.1$) are misclassified as Ia's, while about 90\% of the Ia's are correctly classified
as such. This gives a fraction of false-positives among
the SNe Ia in a photometrically
classified survey to this depth between $(0.3 \cdot 0.2)/(0.3 \cdot 0.2 +0.9 \cdot 0.8)=0.08$
and $(0.3 \cdot 0.5)/(0.3 \cdot 0.5 +0.9 \cdot 0.5)=0.25$.
Alternatively, in deep surveys such as GOODS, about 50\% of
the SNe are CC at $0.3\lesssim z \lesssim 0.9$, and the misclassified CC fraction is
$\sim 35\%$. The SNe Ia are mostly at $0.5\lesssim z \lesssim 1.2$,
$\sim 90\%$ of which are correctly classified, hence the contamination fraction
is slightly higher, of the order of
$(0.35 \cdot 0.5)/(0.35 \cdot 0.5 +0.9 \cdot 0.5)=0.28$.
In the absence of detailed knowledge about the intrinsic SN type
and redshift distributions in a survey, the errors and biases introduced
by misclassification are best estimated using simulations
that test what are the ranges of intrinsic distributions that could
plausibly produce the actually observed, photometrically-typed,
distribution.

\begin{figure}
\epsscale{1}
\plotone{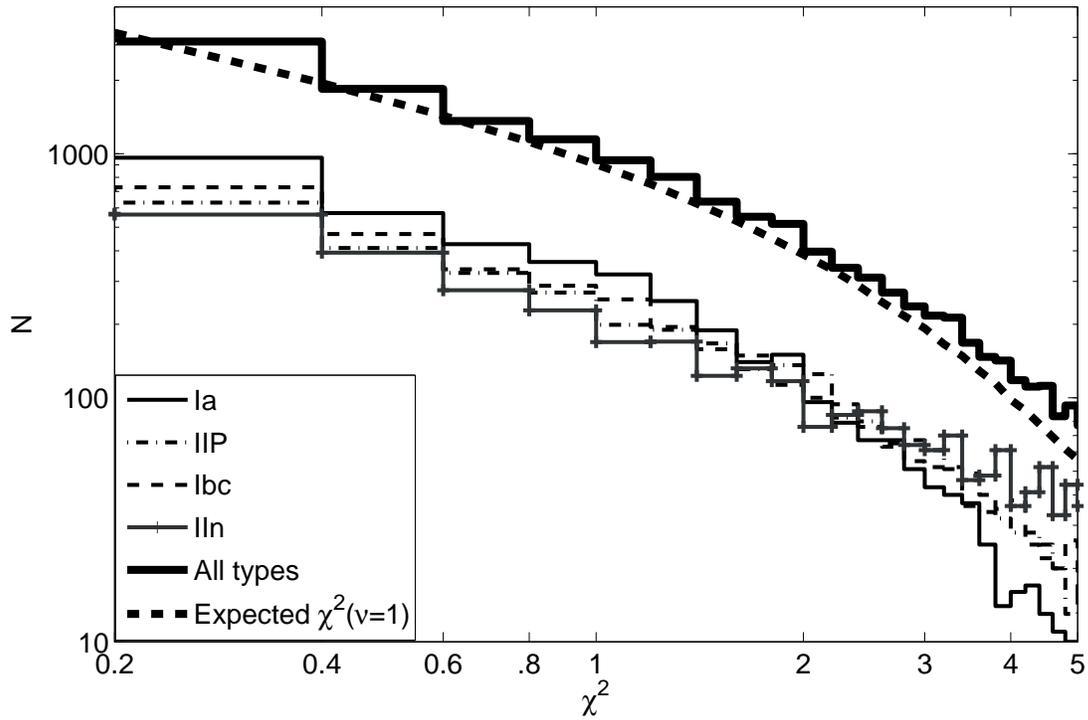}
\caption{Measured $\chi^2$ distributions for the sample of simulated SNe, by type, and for all types combined
(thick solid histogram). The thick dashed curve is the expected distribution for one degree of freedom, normalized to the total number of simulated objects, and matches well the measured curves.\label{chi}}
\end{figure}

We have calculated for each of the simulated SNe the $\chi^2$ value for the best fitting SN template. The $\chi^2$ distribution for each of the SN types can be seen in Figure \ref{chi}. The distribution is similar to the expected $\chi^2$ probability density function with one degree of freedom. Had our model been linear, and its parameters non-degenerate, we would have expected to have zero degrees of freedom, since we fit three parameters (redshift, age, and extinction) to three data points per object (the magnitudes in three bands). Fortunately, our model is non-linear, and has some degeneracies between the parameters, so that useful information can be extracted, as we have shown. From the distribution of
$\chi^2$ values in the simulations, we can derive the $99\%$ confidence interval, which can be used to reject objects which are poorly fit by all of our models. Such objects could be peculiar types of SNe or SN impostors (e.g., AGNs, the major contaminant in SN surveys)
with significantly different colors.
As already noted, in SNLS and GOODS, all the SNe we examined had $\chi^2$ values smaller than the $99\%$ confidence limit of the simulated dataset, meaning there were no true SNe that were misclassified as peculiar/impostor SNe. In our application of SN-ABC to a real ``spectroscopically blind'' SN survey, we can use the $\chi^2$ criterion to assess the contamination by peculiar events and by AGNs. We test this on simulated AGNs in the following section.

Using the same approach used to simulate a SN sample, we have generated 400 mock AGNs using as a template the composite AGN spectrum from \citet{Vanden_berk_01}, and a redshift dependent luminosity function from \citet{BOYLE_00}. We have extrapolated the function to low luminosities based on the evidence presented in \citet{WOLF_03}.
We expect, at least in some cases, the photometric redshift to fail due to the contribution of the AGN to the colors of the host galaxy. As an
approximation of this effect, we generate for each AGN a random host galaxy drawn from a Schechter distribution \citep{schechter_76}. If the AGN luminosity is smaller than the galaxy's, we assume the photo-$z$ works well, and otherwise we generate a completely random photo-$z$. Among the AGNs, 265 are first classified as CC-SNe according to their $P_{Ia}$ values, and the remaining 135 as SNe~Ia. After applying a $\chi^2$ cut, at the 95\% confidence limit, 189 of the CC impostors are rejected, but only eight of the Ia impostors. There is no significant difference between the samples with `good' and `bad' photometric redshifts, so our approximation
for the redshift determination is inconsequential.

Thus, we can reject more than half of the AGNs we simulate. The remaining objects contaminate in about equal fractions the Ia and CC-SN samples. Note that in any search that includes more than two epochs, most of these AGNs will be rejected due to their non-SN-like variability, and survey-specific simulations can be carried out to test for the existence of residual contamination.

\subsection{Number of Photometric Bands}\label{bands}
So far, we have studied the performance of the SN-ABC when classifying SNe with single-epoch photometry in three bands.
We now examine the effect of using more or fewer bands on the classification success rate.

\begin{figure}
\epsscale{0.95}
\plotone{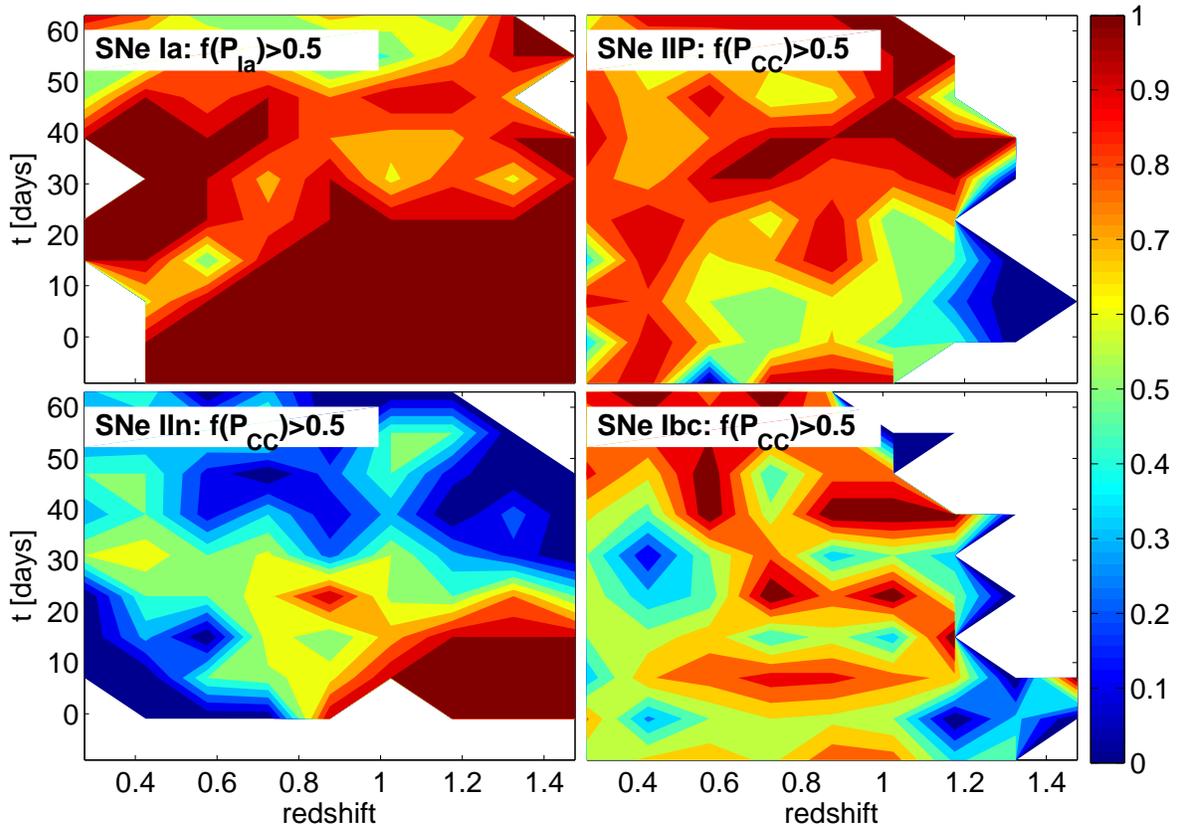}
\caption{Same as Figure \ref{f:MCscss}, but for simulated SNe observed in four bands, rather than three.
Four bands improve the classification success rates of old, low-$z$, SNe~Ia, but have little effect on the success of
CC-SN classification.\label{f:MCscss_4b}}
\end{figure}

We begin with the addition of observations in a fourth, bluer, band, specifically the
HST F475W filter. We repeat the simulations presented in the previous sections, though with fewer SNe (500 of each type), due to computational limitations. Comparing the results shown in Figure \ref{f:MCscss_4b} to those obtained with three bands in Figure \ref{f:MCscss}, one can see that although there is some improvement, it is not striking. The overall success rate for SNe~Ia (top left panel) improves marginally, mainly through a better classification of
1-2 month-old SNe at low redshift.
However, as can be seen in the remaining panels of Figure \ref{f:MCscss_4b}, for the CC-SNe of the different subtypes, the improvement is less significant, and is null for type-IIn SNe.

In order to test whether this improvement remains marginal for real data, we used the SNLS sample which has $g$-band observations for some of the SNe. We repeated the process described in \S\ref{SNLS-Ia}, with the fourth band added, and $\sigma_1=\sigma_2=0.1$. In this case 99\% of the Ia~SNe are correctly classified (only one misclassified), with 74\% having $P_{Ia}>0.9$. While this is better than with three bands, it is not a dramatic improvement.
For $\sigma_1=\sigma_2=0.3$, the correctly classified fraction of SNe~Ia rises from 76\% with three bands to 84\% with four.
The SNe~II-P are only marginally better classified, but the sample with observations in four bands consists of only eight such objects.
We thus conclude that, when the photometric redshifts are reasonably good, observations in a fourth band, for the sake of classification purposes, may be an inefficient use of telescope time.

\begin{figure}
\epsscale{0.95}
\plotone{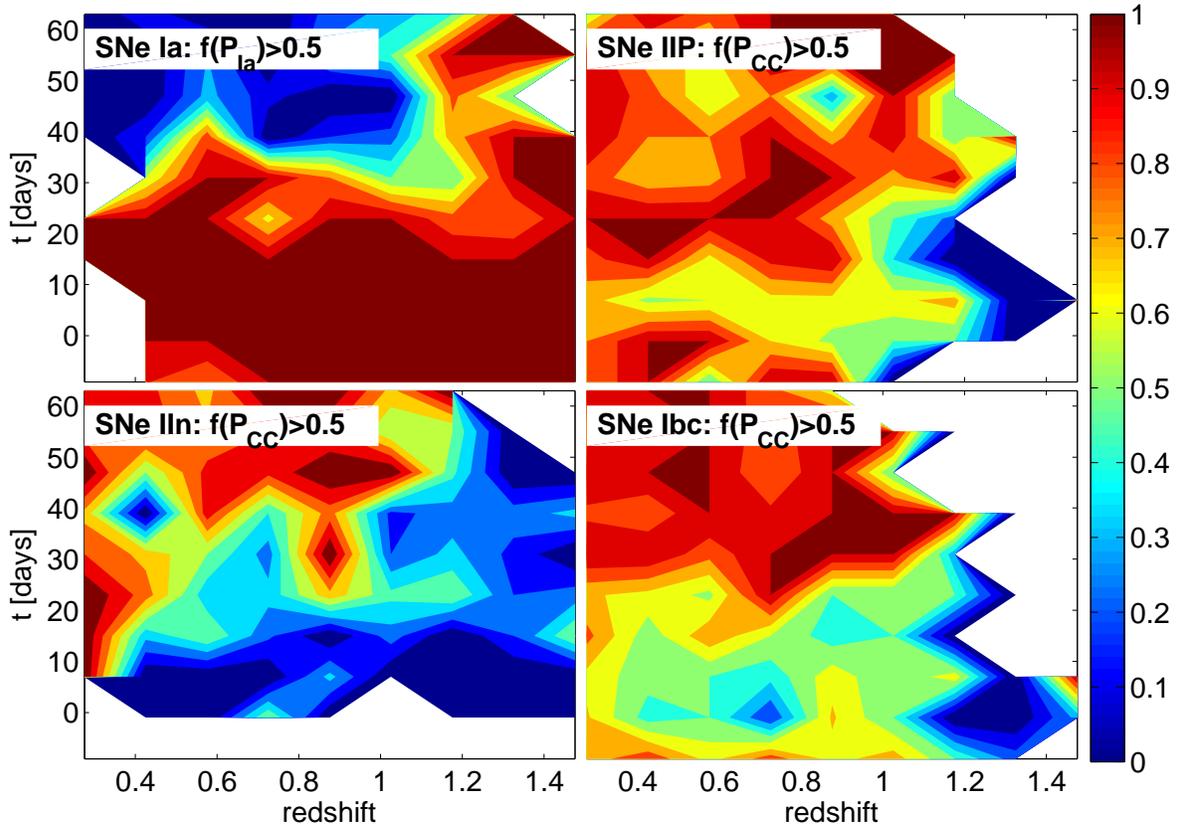}
\caption{Same as Figure \ref{f:MCscss}, but for simulated SNe observed in only two bands (F775W and F850LP).
As expected, SN~Ia classification success is reduced significantly. For CC-SNe, classification success rates
are somewhat higher than with three bands, as explained in \S\ref{bands}.\label{f:MCscss_2b}}
\end{figure}

Next, we examine the success rate of the SN-ABC when using observations only in two bands.
Since the best bands for high-$z$ SN observations are the reddest, we discard the bluest band, and remain with F775W and F850LP.
As expected, with two bands the results, shown in Figure \ref{f:MCscss_2b}, are degraded. Nonetheless, in some regions of parameter space, the success rate of the SN-ABC is non-negligible.
SNe~Ia show the same change as when dropping from four bands to three. The success rates
drop especially at low redshift and for old ages.
On the other hand, surprisingly, the success rate for II-P and Ibc SNe is usually higher with two bands than with three.
This could be understood in the following way. As described in \S\ref{model}, our model for the absolute magnitude of CC-SNe includes a much greater dispersion than that of SNe~Ia, in order to account for their intrinsic diversity. This entails that CC-SNe can populate wider volumes of color-magnitude space, so that when observational constraints are weak (with one or two bands observed) summing the likelihood over all parameter space makes the evidence $E_{CC}$ greater on average than $E_{Ia}$, and thus increases $P_{CC}$ for all objects, regardless of their real type. As a consequence, SN~Ia classification is hindered, while CC-SNe are apparently better classified.
We have tested this explanation by reducing artificially the dispersion of CC-SNe in our model. As a consequence, SN Ia classification improved
dramatically, regardless of the number of observed bands, while CC-SN classification success was correspondingly reduced, being worst with the least number of bands.

\begin{figure}
\epsscale{0.95}
\plotone{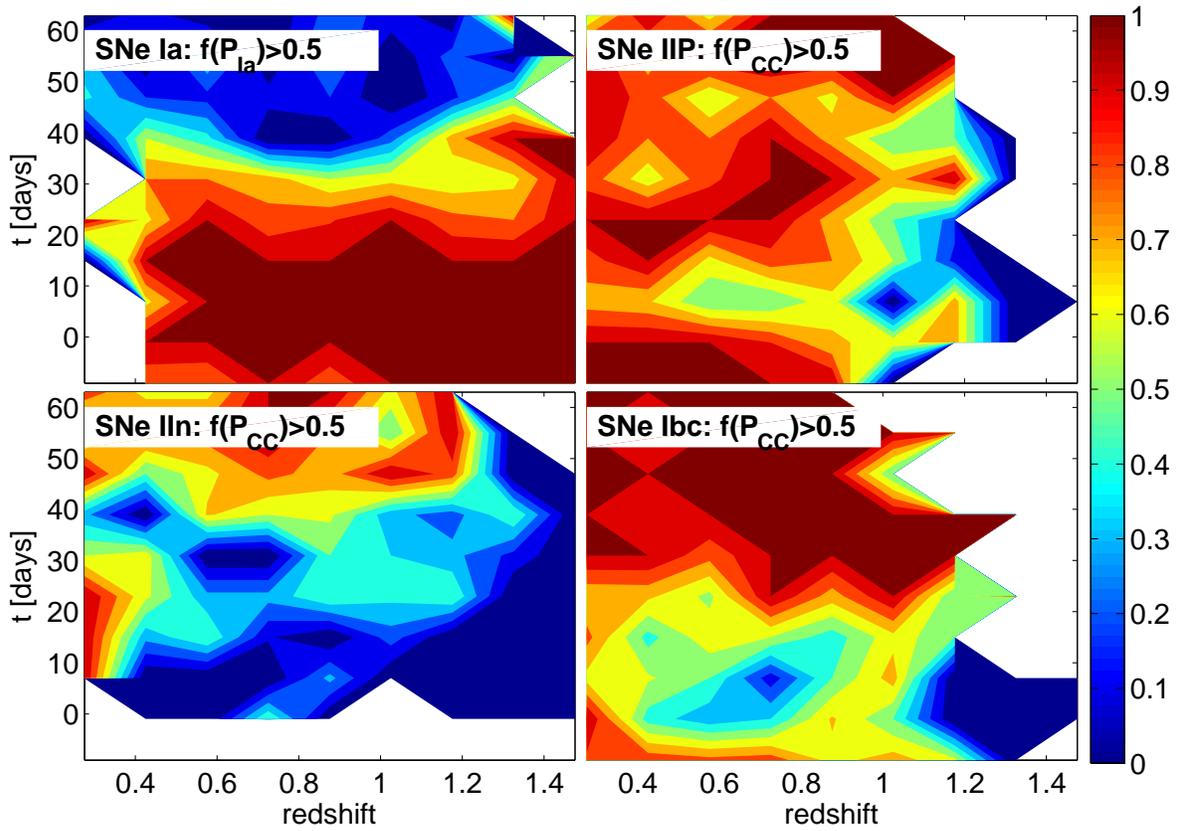}
\caption{Same as Figure \ref{f:MCscss}, but for simulated SNe observed in just one band (F850LP).\label{f:MCscss_1b}}
\end{figure}

Even with only one observed band (the reddest, F850LP), as can be seen in Figure \ref{f:MCscss_1b}, the SN-ABC manages to classify successfully some SNe, e.g. 75-85\% of all the SNe with redshifts $z>0.7$, excluding SNe IIn. This is mainly due to the fact that, at least in our models, SNe~Ia are significantly brighter than most CC-SNe,
which is enough to differentiate (probabilistically) between Ia and CC-SNe when the redshift is constrained.

Finally, AGN rejection, while improving only marginally when adding a fourth band, is weakened considerably when discarding information in one or two bands. This comes from the longer tail of high values that the $\chi^2$
distributions have in these cases. Such a tail increases the 95\% confidence limit by a factor of ten, when compared to the scenarios with three or four bands, and does not allow the rejection of any AGNs.

Thus, while some classification can be done with one or two photometric bands, the biases in the resulting samples are stronger and the success rates are rather low. On the other hand, four bands seem not to be
worth the extra observing time, over observations in three bands, if the sole sole purpose is to classify the SNe based on single-epoch data. This also suggests that, with better sampled light curves, the added value of a blue fourth band is small.

\section{Conclusions}\label{conclusions}

Using the SN-ABC, a Bayesian algorithm, we have studied the feasibility of a probabilistic automatic classification of SNe, based solely on single-epoch multiband photometry and prior redshift information.
While photometrically classified samples are far from competing in precision with spectroscopic samples, they could be an important source of information on the rates and statistical properties of SNe.
We have tested our technique on samples of SNe from the SNLS and GOODS projects that have spectroscopic classification, and on artificial datasets. For most samples and settings our method is promising, with high success rates. When assuming reasonably well-measured photometric redshifts, the SN-ABC correctly classifies
97\% of the type-Ia SNe from SNLS, and 85\% of the type II-P SNe. Similar numbers are achieved for a subset of the GOODS SNe for which there are spectroscopic redshifts. In artificial samples, we have shown that SNe~Ia are best recognized at $z\gtrsim 0.6$, or when near maximum, with success rates of the order of 95\%. Core-collapse SNe are best classified several weeks after maximum, or at $z \lesssim 0.6$
Using the artificial samples, we have shown that, when taking into account the whole zoo of CC-SNe, complications do arise. Specifically, type-Ibc SNe are not as well classified,
and type IIn SNe are usually misclassified.
For surveys where SNe are found young, classification success fractions improve, by a
degree dependent on the SN type and redshift.
The fractional contamination of a single-epoch photometrically selected sample
of SNe~Ia by CC-SNe, could vary between less than 10\% and as much as 30\%.
The errors and biases introduced by misclassification are best estimated using simulations
that test what are the ranges of intrinsic distributions that could
plausibly produce the actually observed, photometrically-typed, distribution.

An additional feature of the SN-ABC is the ability to reject SN ``impostors", mainly AGNs, using $\chi^2$ statistics. In our simulations half of the AGNs we simulated were rejected when using a $\chi^2$ confidence limit of 95\%.
While we have focused on SNe observed in three bands, we have found that a fourth band may be of little additional benefit for the purpose of classification, while with fewer bands some classification is still possible. AGN rejection, however, fails with fewer than three bands.

In all the cases studied, the output values, which we use to classify the SNe, $P_{CC}$ and $P_{Ia}$,
are also indicative of the quality of the fit, and the reliability of the result. The mean values of $P_{CC}$ and $P_{Ia}$ track the success rates of the SN-ABC, and can be used both to define subsamples with lower uncertainties, and to estimate confidence intervals. Methods such as these, and their future refinements, are a prerequisite for pushing the exploration of the SN population to fainter magnitudes and higher redshifts.
A public release version of the SN-ABC is in preparation.

\acknowledgments
We thank the SNLS team, specifically R. Pain and P. Nugent, for access to SNLS data and valuable comments, and S. Leach, E. O. Ofek, and K. Sharon, for useful discussions.
A.G. acknowledges support by NASA through Hubble Fellowship grant \#HST-HF-01158.01-A awarded by STScI, which is operated by AURA, Inc., for NASA, under contract NAS 5-26555.

\clearpage

\clearpage

\clearpage

\clearpage

\clearpage

\clearpage

\clearpage

\clearpage

\clearpage

\end{document}